\documentclass[prl,preprint,showpacs]{revtex4}

\usepackage{amsmath}
\usepackage{amssymb}
\usepackage{amsfonts}
\usepackage{graphicx}
\usepackage{subfigure}
\usepackage{wrapfig}
\usepackage{float}
\floatstyle{boxed}
\newfloat{program}{h}{lop}
\floatname{program}{}

\begin{document}

\title{Thin-thick coexistence behavior of 8CB liquid crystalline films on silicon}

\author{R.~Garcia}
\email{garcia@wpi.edu}

\author{E. Subashi}

\affiliation{Department of Physics, Worcester Polytechnic Institute,
Worcester, MA 01609}

\author{M. Fukuto}

\affiliation{Condensed Matter Physics and Materials Science Department, Brookhaven National Laboratory,
Upton, NY 11973}

\date{\today}

\begin{abstract}
The wetting behavior of thin films of 4'-n-octyl-4-cyanobiphenyl (8CB) on Si is investigated via optical and x-ray reflectivity measurement. An experimental phase diagram is obtained showing a broad thick-thin coexistence region spanning the bulk isotropic-to-nematic ($T_{IN}$) and the nematic-to-smectic-A ($T_{NA}$) temperatures. For Si surfaces with coverages between $47$ and $72\pm3$ nm, reentrant wetting behavior is observed twice as we increase the temperature, with separate coexistence behaviors near $T_{IN}$ and $T_{NA}$. For coverages less than $47$ nm, however, the two coexistence behaviors merge into a single coexistence region. The observed thin-thick coexistence near the second-order NA transition is not anticipated by any previous theory or experiment. Nevertheless, the behavior of the thin and thick phases within the coexistence regions is consistent with this being an equilibrium phenomenon.
\end{abstract}

\pacs{61.30.Hn, 61.30.Pq, 64.70.M-, 64.70.mj, 68.08.Bc, 68.18.Jk, 68.35.Rh, 68.60.-p}
\maketitle
Very thin films of materials on solid surfaces constitute quasi-2D thermodynamic systems, and as such are highly relavent to efforts aimed at understanding how the properties of thermodynamic systems evolve as their size is reduced to ever smaller dimensions \cite{wetting,ziherl}.  Recently, thin liquid crystalline films of 4'-n-pentyl-4-cyanobiphenyl (5CB) were discovered to exhibit a thin-thick coexistence region (CR) on a silicon surface \cite{veffexpt}. Just as dense (liquid) and rare (gas) phases coexist over a range of temperatures for a fixed number of fluid molecules in a container, for a fixed number (coverage) of 5CB molecules on Si, there is a temperature range just below the bulk first-order isotropic-to-nematic (IN) transition temperature $T_{IN}$  where thick-film and thin-film phases coexist on the surface. The percent area coverage by each phase is found to be consistent with a lever rule that conserves the total average surface coverage. Two features of the thick-thin coexistence have favored its interpretion as an instance of the bulk transition being continued down to 2D. First, the temperature-width of CR shrinks with increasing coverage, so that it becomes indistinguishable from the bulk transition for coverages greater than 100 nm. Second, optical properties of the coexisting phases appear to be consistent with the thin-film phase being isotropic and the thick-film being nematic \cite{veffexpt}. 

Nevertheless, the 5CB experiments and their interpretation have proved highly controversial. One criticism is that the films visibly dewet from the substrate over a period of 1-2 days, suggesting that the coexistence may be non-equilibrium in nature \cite{ziherl}.  An equilibrium model by van Effenterre et al. \cite{veff} attempts to explain the observed $1/d^2$ shape of the coexistence region boundary purely in terms of mean-field elastic forces due to director gradients within the film. However, it has been criticized because it makes two potentially unphysical assumptions: (1) that an equilibrium isotropic phase could exist several degrees below $T_{IN}$, and (2) that a $0.5$ radian gradient in the director is present in the nematic film phase, even though for this regime of film thickness the interplay of anchoring energies favors a uniform director field in the film \cite{ziherl}. 

The 4'-n-octyl-4-cyanobiphenyl (8CB) molecule, which is slightly longer than 5CB and forms cybotactic dimers, has been shown to stably wet Si, spreading on the surface in a layer-by-layer fashion \cite{8CBspreading}. 8CB also significantly differs from 5CB by possessing both a weakly first-order IN transition at $T_{IN}=40.5^{\circ}$C and a second-order nematic-to-smectic (NA) transition at $T_{NA}=33.4 ^{\circ}$C \cite{criticalNA,8CBtemp}. The controversial van Effenterre model \cite{veff}, which does not take into account such smectic order, could not be expected to apply to 8CB films, except as an approximation near $T_{IN}$. Furthermore, the observed persistence of smectic order at the vapor and substrate interfaces up to $41^{\circ}$C \cite{8CBxray,8CBspreading} makes the significant director gradients required by the model in the middle of the film even harder to imagine.  Near $T_{NA}$, limited data and theory available suggest that fluctuation-induced forces would be dominated by an elastic force which oscillates with a period of 3.2 nm (smectic spacing) as the film thickness increases \cite{NAstructuralforces}. The question, however, of how the film will behave in the vicinity of the second-order NA phase transition \cite{criticalNA}, remains largely uninvestigated both theoretically and experimentally \cite{ziherl,NAstructuralforces}.

In this letter, we present high-resolution optical and x-ray reflectivity measurements of 8CB films which demonstrate the same coexistence behavior near $T_{IN}$ as seen in 5CB alongside a new unexplained effect near $T_{NA}$. The substrates for our experiments consist of $2 \times 2$ cm squares of B-doped (10-20 $\Omega/\square$) Silicon (100) with a 2.0 nm-thick layer of native oxide. RCA cleaning procedures for removing organic and ionic residues from the substrates \cite{RCA} had no discernible effect on the observed phenomena. Thus, to minimize sample handling and accumulation of dust particles on the surface, substrates are mostly used directly as obtained from the manufacturer. 

The 8CB films are prepared by spin coating from a dilute solution of 8CB in chloroform containing 1 \% ethanol as a stabilizer. By varying the 8CB concentration (between 1:150 and 1:460 by volume of liquid crystal to chloroform) and setting the acceleration on a programmable Laurel spin coater between $4.5 \times 10^{2}$ and $2.2 \times 10^{4}$ rpm/s, films were prepared with thicknesses between 20 and 120 nm thick, as determined by the wavevector period $\Delta q = 2\pi/d$ of the observed interference fringes in the measured x-ray reflectivity (e.g., Fig. 3). The film thicknesses were insensitive to the final rpm value, the precise size of the substrate, and ambient humidity (varying 10-50\% during the experiments). The 8CB films appear completely stable, not changing for over 10 days under constant room temperature conditions.

For each experimental run, a spin-coated 8CB film of a desired coverage (uniform thickness at room temperature) is placed inside a temperature-regulated cell, where micrographs are taken every few minutes as the temperature is increased at 0.025 $^{\circ}$C/min. The experimental cell, machined out of aluminum, is fitted with Pt thermometers calibrated to within an accuracy of $\pm 0.2 ^{\circ}$C, and has $\sim 2 \times 2$ cm indium-tin-oxide coated windows on their upper side for optical access. As the temperature is increased, it is controlled to within $\pm0.02 ^{\circ}$C. Micrographs of the films are taken using a inspection microscope and computer-interfaced CCD camera.  As we scan the temperature, the thin-thick coexistence region (CR) is signalled by sharp non-uniformity in the color of reflected light from the film. 

Figure~1 shows the temperature-thickness ($T$-$d$) phase diagram obtained by analyzing micrographs from the experiment. The first striking feature of Fig.~1 is the dashed dewetting line DW, which is the uppermost line of the diagram, given approximately by $T_{DW}= (44.1\pm0.3)\text{$^{\circ}$C} - {(3600 \pm300)\text{nm$^2$ $^{\circ}$C}}/{d^2} .$
Below this line, 8CB appears completely stable on the surface, but above it, the 8CB spontaneously dewetts from the Si surface over a period of minutes. The large-thickness limit of 44.1 $^{\circ}$C for this line agrees with the dewetting temperature of $42\pm5$ $^{\circ}$C previously observed for 8CB droplets on Si \cite{8CBspreading}. Although the $1/d^2$ form is consistent with structural and surface tension forces \cite{veff,veffexpt,8CBnylon}, there is, at present, no specific quantitative prediction for comparison. 

The second and main result is the observation of the crescent-shaped, thin-thick coexistence region CR that is enclosed by solid lines in Fig.~1. This diagram not only confirms thin-thick coexistence near the bulk $T_{IN}$ (already seen in 5CB), but it also shows a completely new feature, namely, a distinct coexistence region near the bulk $T_{NA}$. Characteristic micrographs of the film in various regions in the phase diagram are shown in Fig.~2.  Outside CR and below DW, the stable state consists of a uniform thickess film covering the entire surface. Inside CR, two distinct phases, thick and thin, are observed. As shown in the inset of Fig.~1, the observed percentage area covered by each phase at a given $T$ agrees roughly with the percentage calculated from the two $d$ values given for the boundary of CR for that $T$ and the requirement that the total coverage is constant on the surface.

The scatter in the phase diagram is due in part to a $\pm 7$\% uncertainty in $d$ inherent in our film preparation technique and in part due to the difficulty in visually identifying the precise temperature for the onset or disappearance of islands. Enhanced scatter is observed in the nematic region just above $T_{NA}$, where theory predicts there should be stresses that oscillate as $d$ increases with the smectic spacing of 3.2 nm \cite{NAstructuralforces}. Although it is tempting to try to relate this scatter to such forces, the observed scatter does not appear to repeat every 3.2 nm. Within the scatter shown, the features of the phase diagram are reproducible on heating and cooling if we remain below DW. Although increasing the temperature scan rate by a factor of two had an observable effect on island distribution within CR, the percent area covered by each phase was unaffected and the effect on the determination of CR and DW boundaries was less than the scatter in the phase diagram. When cooling the films slowly, the films get stuck in whatever island pattern is present at the bottom of the CR near $T_{NA}$, a non-equilibrium artifact known as the ``surface memory effect'' in liquid crystals \cite{clark}. To avoid this effect, it is necessary either to cool extremely slowly or to rapidly quench the film through the CR. To circumvent these complications, almost all of our experimental runs were done on heating.

The existence and shape of the CR near $T_{IN}$ are similar to the results obtained previously for 5CB \cite{veff}. For $T > 37.0 ^{\circ}$C, the CR boundaries can be well described, as shown in Fig. 1 (solid lines), by $ T(d) = T_{IN}  - \frac{C}{d^2}$ where for the upper and lower boundaries, respectively, $C = (1730 \pm130) $ and $(5700 \pm900) \text{nm$^2$ $^{\circ}$C}$. These values are in contrast to $C = (760 \pm40)$ and $(5100 \pm 500)$ nm$^2$ $^{\circ}$C, respectively, for 5CB on Si ~\cite{veff}. 
By contrast, the boundaries of the CR \textit{below} $37.0 ^{\circ}$C in Fig.~1 are drawn as guides to the eye and not fits to any theory. The rounded shape near $T_{NA}$ has been extrapolated from the percent areal fractions of the thick and thin phases within the CR by assuming that the thickness of the thin phase is given by the left boundary of the CR. 

Due to the shape of the CR, the behavior as we increase $T$ depends on the surface coverage. For coverages greater than $70\pm 3$ nm, thin-thick coexistence is only observed near $T_{IN}$ and not near $T_{NA}$. For these coverages, micrographs show changes in film roughness 0-2 $^{\circ}$C below the bulk $T_{NA}$ that may be indicative of the NA transition in the film, which may terminate at or merge with the rightmost portion of CR. Further study, however, is needed to measure $T_{NA}(d)$ directly. As $T$ is increased, small areas of thin film phase first nucleate upon reaching the bottom of CR just below $T_{IN}$; thereafter, the percent area covered by the thin phase increases approximately linearly with $T$, until at the top of the CR the thin phase covers the entire surface. Consistent with the constraint of constant surface coverage, x-ray reflectivity shows that the thickness of the thin phase at the top of CR is within 1 nm of the thickness of the thick-film phase at the bottom.

For coverages between $47$ and $70 \pm 3$ nm, with increasing $T$ the CR is crossed twice: first near $T_{NA}$ and second near $T_{IN}$.  For a coverage of 60 nm, thin-film islands first nucleate at 31.8 $^{\circ}$C. These islands expand in area and multiply up to 32.8 $^{\circ}$C, above which they decrease in number and area, disappearing completely near 34.5 $^{\circ}$C. Thereafter the behavior is identical to what is seen for coverages above $70$ nm. The film remains completely uniform until it re-enters and re-exits CR just below $T_{IN}$. Thin film islands reappear near 37.8 $^{\circ}$C, then expand in area until they cover the entire surface at 39.5 $^{\circ}$C. 

Finally, for coverages between $20$ and $47\pm 3$ nm, only one continuous coexistence is observed, beginning just below $T_{NA}$ and ending just below $T_{IN}$. In this case, islands of thick (not thin) film are nucleated at the bottom of CR, due to entering CR on the left, thin-film side (Figs. 1-2). For a coverage of 27 nm, islands of the thick-film phase first form at 33 $^{\circ}$C, then grow and multiply until they cover about 30\% of the surface at 36 $^{\circ}$C. Increasing $T$ further, these islands diminish, finally disappearing at 37.8 $^{\circ}$C, leaving a uniform film. 

In this study, x-ray reflectivity, performed using the liquid surface diffractometer at Beamline X-22B of the National Synchrotron Light Source (NSLS), was used primarily to obtain a film thickness scale for our measurements. However, reflectivity scans also offer insight into the structure of films at various points on the phase diagram. In Fig.~3, x-ray reflectivity scans are shown for a film initially 66 nm thick at 26.0, 33.3, 35.1 and 38.4 $^{\circ}$C. X-ray and optical measurement of the films was peformed simultaneously. Each of these points was located outside CR, where the film appeared completely uniform. For measurements at NSLS, the CR boundaries were shifted downwards by 1 $^{\circ}$C, possibly an effect due to dust on the surface. At 26 $^{\circ}$C, the amplitude of reflectivity oscillations increases with wave vector transfer $q (>1.2\text{nm$^{-1}$})$ building to a prominent Bragg peak at $q_o = 2.0$ nm$^{-1}$. The thickness of each smectic layer in the film is $a=2\pi/q_o= 3.2$ nm, and, based on the separation of $0.2$ nm$^{-1}$ between the minima on either side of $q_o$, the total thickness of the smectic layers is $Na \sim 4\pi/(0.2 \text{nm}^{-1})=63$ nm $\sim d$, implying smectic ordering throughout the film. In the uniform film region above $T_{NA}$ at 35 $^{\circ}$C, the Bragg peak is diminished and broadened, and the reflectivity oscillations decay with $q$, consistent with the near absence of smectic ordering except possibly at the film's solid and vapor interfaces \cite{8CBxray}.  At 38.4 $^{\circ}$C, above CR near $T_{IN}$, the Bragg-peak-like feature centered at $q_o$ is further diminished. 

In conclusion, 8CB films on Si are found to exhibit thin-thick coexistence behavior near not only the first-order IN but also near the second-order NA transition. Despite the fact that 8CB is quite different from 5CB in many respects, experimentally it is found to behave similarly near $T_{IN}$, with the shape of the thin-thick coexistence region being consistent with a $\Delta T \sim 1/d^2$ law. Near $T_{NA}$, 8CB's SmA nature, however, leads to an unexpected unique proturberance in the shape of the coexistence region, which has not been explained. Unlike previously investigated liquid crystal systems, the as-prepared 8CB film at room temperature is completely stable for all film thicknesses studied, so long as the system remains below a dewetting line $T_{DW}(d)$ that lies above the coexistence region. More detailed measurement and analysis are necessary in order to determine the smectic, nematic or isotropic nature of the films in the various regions of the phase diagram, as well as to better discern the shape of the coexistence region near $T_{NA}$. We hope that these results will stimulate efforts to theoretically model the nature of sructural forces in the vicinity of the second-order NA transition in 8CB films. 

We thank G. Iannacchione, B. Ocko, P. Pershan, and R. Pindak for useful discussions and comments. This research was supported by PRF Grant No. 45840-G5. The work at Brookhaven National Laboratory and National Synchrotron Light Source were supported by the U. S. Department of Energy, under contract no. DE-AC02-98CH10886.


\begin{figure}[ht]
\vskip 0.1 cm 
\includegraphics[angle=-90,width=82mm]{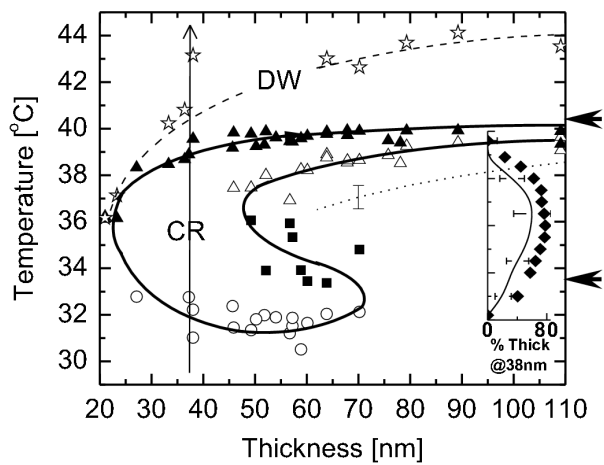} 
\vskip 0.1 cm \noindent
\caption{Temperature-thickness phase diagram for 8CB on Si. As the temperature is increased, open symbols indicate when islands first appear on the surface, while closed symbols indicate when islands disappear leaving a uniform film.  Above the dashed dewetting line DW, the film spontaneously dewetts from the Si surface, forming droplets. Below DW, the film stably wets the surface. Inside the thick-thin coexistence region CR, at a given temperature, two distinct, stable film thicknesses are observed on the surface that appear to correspond to the thicknesses indicated by the left and right boundaries of CR (solid lines) at that temperature. The arrows at the right indicate the bulk $T_{NA}=33.4 ^{\circ}$C and $T_{IN}=40.5^{\circ}$C. The dotted line is $T_{IN}(d)$ for confined 8CB extrapolated from measurement/model of Ref. \cite{8CBnylon} for $d>170$ nm. The vertical arrow at 38 nm shows a typical constant coverage path through the phase diagram.  The inset shows observed percent area covered by thick phase (diamond) versus a computation  (solid line) based on the shape of CR and the constant coverage path at 38 nm. \label{fig1} }
\end{figure}

\begin{figure}[ht]
\vskip 0.1 cm 
\includegraphics[angle=0,width=88mm]{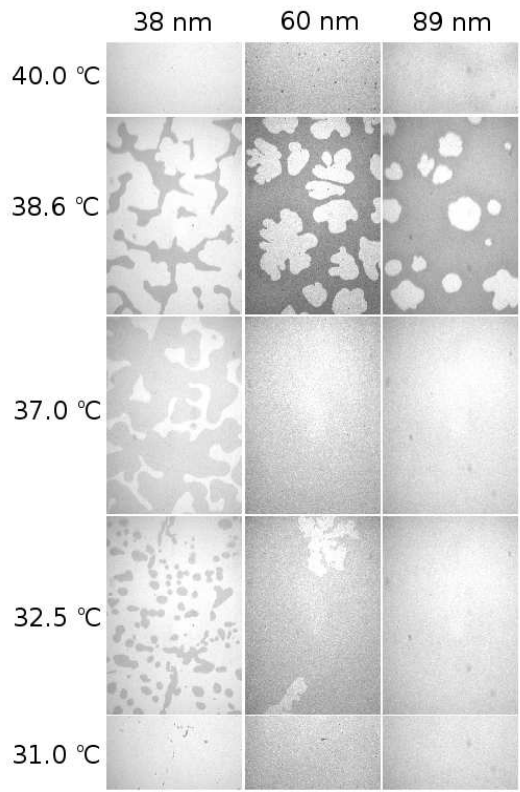} 
\vskip 0.1 cm \noindent
\caption{Typical micrographs of 38, 60 and 89 nm thick films as $T$ is increased by 0.025 $^{\circ}$C/min. Darker shade of grey indicates thicker film. The micrograph displayed for the 89 nm film in the row labeled 38.6 $^{\circ}$C was actually taken at 39.5 $^{\circ}$C. Note that the 60 nm film exhibits coexistence behavior twice with increasing temperature. Each image is 1mm across. \label{fig2} }
\end{figure}

\begin{figure}[ht]
\vskip 0.1 cm 
\includegraphics[angle=-90,width=82mm]{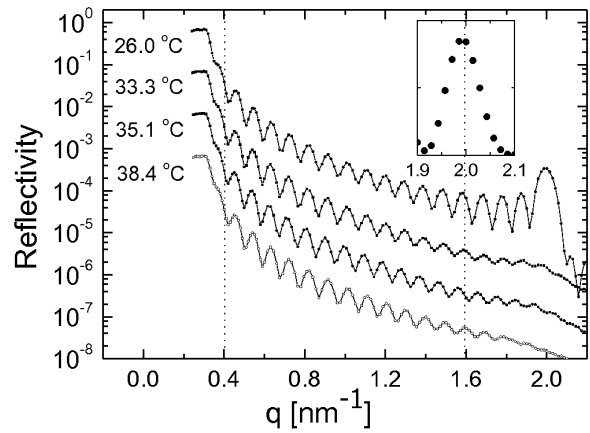} 
\vskip 0.1 cm \noindent
\caption{ X-ray reflectivity versus wave vector transfer $q$ for a film initially $66\pm2$ nm thick (from the wavevector period) at 26.0, 33.3, 35.1 and 38.4 $^{\circ}$C. For clarity, each higher temperature curve is shifted vertically downwards by a factor of 10. The vertical lines are drawn as a guide to the eye. The inset shows the Bragg peak at $q_o = 2.0$ nm$^{-1}$ and the two adjacent minima on either side for the 26 $^{\circ}$C data on a linear scale.\label{fig3} }
\end{figure}

\end{document}